\def\be{\begin{equation}}
\def\bea{\begin{eqnarray}}
\def\ee{\end{equation}}
\def\eea{\end{eqnarray}}
\def\dd{\displaystyle}
\def\nn{\nonumber}
\documentclass[epj]{svjour}
\usepackage{graphics}
\begin{document}
\title{Extra Dimensions in Particle Physics}
\author{Ferruccio Feruglio\inst{1}% etc
}                     % Do not remove
\institute{Universit\`a di Padova \and I.N.F.N. sezione di Padova}
\abstract{Current problems in particle physics are reviewed
from the viewpoint of theories possessing extra spatial dimensions.
%
%\PACS{
%      {PACS-key}{discribing text of that key}   \and
%      {PACS-key}{discribing text of that key}
%     } % end of PACS codes
} %end of abstract
\maketitle
\section{Introduction}
\label{intro}
Today extra dimensions (ED) represent more a general framework
where several aspects of particle physics have been reconsidered,
rather than a unique and specific proposal for a coherent
description of the fundamental interactions.
The first motivation for the appearance of ED,
namely the quest for unification of gravity
and the other interactions \cite{kk}, is still valid today.
If we strictly adhere to this project, then at present the 
only viable candidate for a unified description of all 
interactions is string theory, which naturally requires ED.
The idea of ED that we have in mind today has been deeply
influenced by the developments in string theory: 
compactifications leading to a chiral fermion spectrum, 
localization of gauge and matter degrees of freedom
on subspaces of the ED,
relation between the topological properties of the compact 
space and the number of fermion 
families, localization of states around special points of
the compact space and hierarchical Yukawa couplings,
just to mention few examples.
Remarkable theoretical progresses have also been obtained 
by developing models in field theory. For instance, in
this context very fruitful tools for supersymmetry
(SUSY) and gauge symmetry breaking have been developed,
such as the Scherk-Schwarz \cite{ss} and the Hosotani \cite{hos} mechanisms. 
Also the physical properties of compactifications with
non-factorizable space-time metric have been neatly worked out \cite{rs}. 
The conceptual and mathematical richness offered 
by these developments makes it possible to reconsider
several specific problems that have not received a satisfactory
answer in the four-dimensional context:

\begin{itemize}
\item[$\bullet$]
{\bf Hierarchy problem}: extreme weakness of gravity in comparison to the other
interactions; gap between the electroweak scale and the Planck scale 
$M_{Pl}$.
\item[$\bullet$]
{\bf Little hierarchy problem}: possible gap between the Higgs mass and 
the electroweak symmetry breaking scale.
\item[$\bullet$]
{\bf Problems of conventional Grand Unified Theories (GUTs)}:
doublet-triplet splitting problem, proton lifetime, mass relations.
\item[$\bullet$]
{\bf Flavour problem}: the architecture underlying the observed
hierarchy of fermion masses and mixing angles.
\item[$\bullet$]
{\bf Cosmological constant problem}:
small curvature of the observed space-time
and its relation to the dynamics of particle interactions.
\end{itemize}
In this talk I will review the viewpoint on these problems offered
by theories with ED, stressing the most recent developments in the
field.

\section{Hierarchy Problem}

\subsection{Large Extra Dimensions}

\begin{figure}[h!]
\resizebox{0.40\textwidth}{!}{%
  \includegraphics{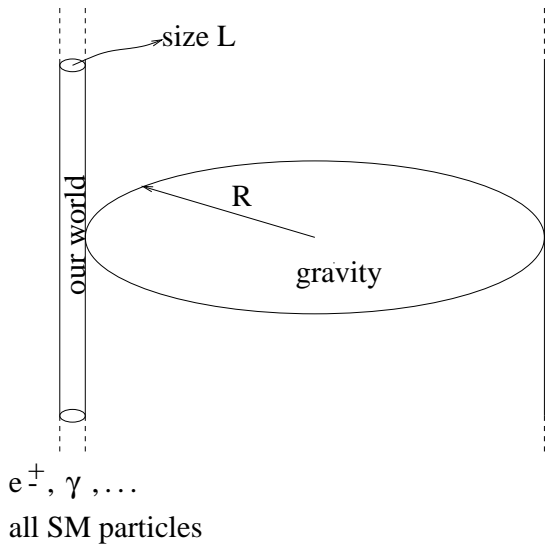}
}
\vskip 0.3cm
\caption{Pictorial view of the generic set-up considered in
this review. Gravity has access to the full space-time
characterized by extra spatial dimensions of typical size $R$.
SM particles are localized in subspaces of the full
space-time, whose typical extension in the extra space is $L<<R$.
When discussing LED, we will set $L=0$.}
\label{fig1}     
\end{figure}

The hierarchy problem can be reformulated in the context of
large extra dimensions (LED) \cite{add}. In the LED scenario there is only one 
fundamental
energy scale for particle interactions: the TeV scale. Gravity describes
the geometry of a $D=4+\delta$ dimensional space-time where $\delta$
dimensions are compactified, for instance, on an isotropic torus $T^\delta$
of radius $R$. The D-dimensional Planck mass $M_D$ is of the order 
1 TeV.
All the other degrees of freedom of the standard model (SM)
are assumed to live on a four-dimensional  subspace,
usually called brane, of the full D-dimensional space-time
(see fig. \ref{fig1}).
The gravitational potential $V(r)$ between two massive particles
at a distance $r$ has two regimes. For $r\ll R$, the lines of the
gravitational field extend isotropically in all directions
and 
\be 
V(r)\propto \frac{1}{M_D^{2+\delta}}\frac{1}{r^{1+\delta}}~~~,
\ee
whereas for $r\gg R$, the lines are squeezed along
the usual four dimensions:
\be 
V(r)\propto \frac{1}{M_D^{2+\delta}V_\delta}\frac{1}{r}~~~,
\ee
where $V_\delta$ is the volume of the compact space. As a consequence
the four-dimensional Planck mass $M_{Pl}$ is given by the relation\footnote{Several conventions exist in the literature to
introduce the fundamental scale of gravity $M_D$. In this review  
the relation (\ref{dtop}) defines the (reduced) D-dimensional
Planck mass $M_D$ in terms of the reduced four-dimensional Planck mass
$M_{Pl}=2.4\times 10^{18}$ GeV.}:
\be
\left(\frac{M_{Pl}}{M_D}\right)^2=M_D^\delta V_\delta~~~.
\label{dtop}
\ee
Therefore a huge hierarchy between $M_{Pl}$ and $M_D\approx 1$ TeV  
is possible if the volume $V_\delta$ of the extra space is much 
larger than naively expected, $M_D^{-\delta}$.
In units where $M_D=1$ we have
\be
\frac{1}{M_{Pl}}=\frac{1}{\sqrt{V_\delta}}~~~,
\ee
indicating that four-dimensional gravity is weak because the four-dimensional graviton
wave function is diluted in a big extra space.
The $1/\sqrt{V_\delta}$ suppression of the gravitational constant
can also be understood as the normalization factor 
of the wave function for the graviton zero mode.
Solving the hierarchy problem now requires explaining why
$V_\delta\gg M_D^{-\delta}$. 
This is a dynamical problem, since in higher-dimensional
theories of gravity the volume $V_\delta$ is the vacuum expectation 
value (VEV) of a field. For instance in a five-dimensional theory,
the compactification radius $R$ is determined by the 
VEV of the radion field $r(x)$, the (zero-mode of the) 
55 component of the full space-time metric:
\be
\langle r(x)\rangle=R M_D~~~~~,
\ee
in close analogy to the Fermi scale, determined in the electroweak
theory by the VEV of the Higgs field.
In a realistic theory of gravity like string theory, the solution to
this dynamical problem requires finding
the minimum of an energy functional depending simultaneously
upon many moduli, a rather formidable task. 

In table 1, we read the radius $R$ and the compactification scale 
$R^{-1}$ as a function of the number $\delta$ of ED,
assuming $M_D=1$ TeV, an isotropic toroidal compactification
and a flat background metric. Of special interest is the case
$\delta=2$, which predicts deviations from the Newton's law
at distances around 1 mm. For $\delta>2$ such deviations would
occur at much smaller length scales, outside the range of the
present experimental possibilities. For the values of $\delta$
quoted in table 1, the compactification scale is quite small and
the Kaluza-Klein (KK) modes of gravitons can be 
produced both at colliders and in processes of astrophysical and/or
cosmological relevance.
\\[0.1cm]
{\begin{center}
\begin{tabular}{|c|c|c|} 
\hline   
& &\\
$\delta$ & $R$ & $R^{-1}$\\
& &\\
\hline
& &\\                         
1 & $10^8$ Km & $10^{-18}$ eV\\
& &\\
\hline
& &\\                         
2 & $0.1$ mm & $10^{-3}$ eV\\
& &\\
\hline
& &\\                         
3 & $10^{-6}$ mm & $100$ eV\\
& &\\
\hline
& &\\                         
... & ...  & ...\\
& &\\
\hline
& &\\                         
7 & $10^{-12}$ mm & $100$ MeV\\
& &\\
\hline
\end{tabular} 
\end{center}}
\vspace{3mm}
Table 1: Compactification radius $R$ and 
compactification scale $R^{-1}$ as a function of the 
number $\delta$ of ED, for $M_D=1$ TeV.

\subsubsection{Deviations from Newton's law}
Modifications of the laws of gravity
at small distances are currently under intense investigation.
In experimental searches, deviations from the Newton's law are 
parametrized by the modified gravitational potential
\be
V(r)=-\frac{G_N m}{r}
\left(1+\alpha~ e^{-\dd\frac{r}{\lambda}}\right)~~~,
\ee
($G_N=1/(8\pi M_{Pl}^2)$ is the Newton constant)
in terms of the relative strength $\alpha$ and the range $\lambda$
of the additional contribution.
Precisely this kind of deviation is predicted by LED for $r\ge \lambda$.
The range coincides with the wavelength of the first KK
graviton mode, $\lambda=2 \pi R$, while the relative strength
is given by the degeneracy of the first KK level: $\alpha=2\delta$.
At present the best sensitivity in the range $\lambda\approx 100~\mu$m
has been attained by the torsion pendulum realized by the E\"ot-Wash 
group. For $\delta=2$ they obtained the limit \cite{new1}
\be
\lambda<150~ {\rm \mu m}~~~ {\rm (95\%~ CL)} ~~~,
\ee
already in the range where deviations are expected
if the $M_D$ is close to 1 TeV.
Future improvements represent a real challenge from
the experimental viewpoint, but their impact could be
extremely important, since deviations around the currently
explored range are also expected in other theoretically
motivated scenarios, as we shall see later on. 
The next planned experiments aim to reach a sensitivity
on $\lambda$ in the range (30-50) $\mu$m for $\alpha=4$
\cite{new2}, thus probing $M_D$ up to $3\div 4$ TeV.

\subsubsection{Neutrino Masses}
LED provide a nice explanation of the smallness of neutrino masses \cite{ednu}.
If right-handed neutrinos $\nu_s$ ($s\equiv$singlet under the 
SM gauge interactions) exist,
at variance with the charged fermion fields, 
they are allowed to live in the bulk of a LED. 
In this case, just as for the graviton, their four-dimensional modes 
in the Fourier
expansion carry a suppression factor $1/\sqrt{V_\delta}$:
\be
\nu_s(x,y)=\frac{\nu_s^{(0)}(x)}{\sqrt{V_\delta}}+...
\label{fourier}
\ee
By taking into account the relation (\ref{dtop}), the Dirac
neutrino masses originating from the Yukawa coupling with the 
Higgs doublet are given by:
\be
{\cal L}_{Yuk}=\frac{y_\nu v}{\sqrt{2}} \left(\frac{M_D}{M_{Pl}}\right)
\nu_a(x)\nu_s^{(0)}(x)+...
\ee
where $\nu_a(x)$ are the active four-dimensional neutrinos.
The resulting Dirac neutrino masses are much smaller than
the charged fermion masses and the observed smallness of neutrino masses
is explained, if there are no additional contributions. 
The latter might arise from dimension five operators associated to 
the violation of the lepton number. 
In the absence of a sufficiently large fundamental
scale, new mechanisms should be introduced to guarantee the 
desired suppression of these operators \cite{edbl,split}.
\begin{figure}[h!]
\resizebox{0.3\textwidth}{!}{%
  \includegraphics{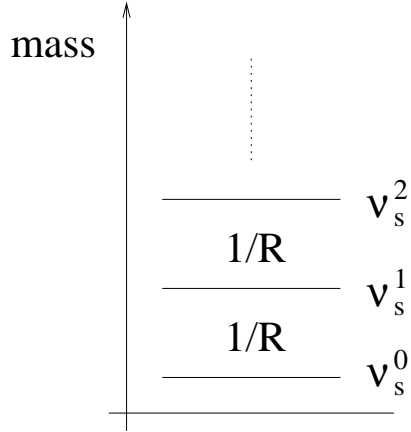}
}
\vskip 0.3cm
\caption{Mass spectrum of right-handed neutrinos $\nu_s^{(n)}$.}
\label{fignu}     
\end{figure}
Testing this idea and detecting the higher-dimensional origin of 
neutrino masses would represent a clean signature of the LED
scenario. Experimentally, this is only possible if some $\nu_s^{(n)}$ 
(see fig. \ref{fignu})
is sufficiently light, $1/R\le \sqrt{\Delta m_{atm,sol}^2}
\approx 0.01$ eV, which is not unconceivable if there is 
one dominantly large ED with $R\approx 0.02$ mm.
In such a case few $\nu_s^{(n)}$ levels may
take part in neutrino oscillations and produce observable
effects.
Unfortunately present data disfavour this exciting possibility.
First of all, there are clear indications in favour
of oscillations among active neutrinos, both in the solar
and in the atmospheric sectors \cite{mura}. Moreover SN1987A
excludes the large mixing angle between active and sterile
neutrinos that would be needed to reproduce, for instance,
the solar data in this scenario \cite{sn1987}. 
Therefore the effects of $\nu_s^{(n)}$ on neutrino oscillations
are sub-dominant, if present at all.
Indeed, if the KK levels $\nu_s^{(n)}$
are much heavier than the mass scale relevant for neutrino 
oscillations, they decouple from the low-energy theory
and the higher-dimensional origin of neutrino masses becomes
undetectable. 

\subsubsection{Signals at colliders}
Experimental signatures of LED at present and future colliders
are well understood by now \cite{grw} and an intense experimental
search is currently under way. The existence of light 
KK graviton modes leads to two kinds of effects. The first one
is the direct production of KK gravitons in association with a photon
or a jet, giving rise to a signal characterized by missing
energy (or transverse energy) plus a single photon or a jet.
The cross section for the production of a single KK mode is depleted
by the four-dimensional gravitational coupling, $1/M_{Pl}^2$, but
the lightness of each individual graviton mode makes it possible
to sum over a large number of indistinguishable final states
and the cross section for the expected signal scales as:
\be
\sigma\approx \frac{E^\delta}{M_D^{\delta+2}}
\ee
in terms of the available center of mass energy $E$.
The present lower bound on $M_D$, listed in table 2,
are dominated by the searches for $e^+ e^-\to\gamma +
\not\!\!{E}$ at LEP and $p {\overline p}\to\gamma +
\not\!\!{E}_{T}$ at Tevatron. For $\delta=2,3,4$ the limits
from LEP are slightly better than those from Tevatron,
the opposite occurring for $\delta=5,6$.  
Recently, comparable limits have also been obtained at Tevatron
\cite{mtej},
by looking for final states with missing transverse energy
and one or two high-energy jets.
\\[0.1cm]
{\begin{center}
\begin{tabular}{|c|c|c|c|c|c|} 
\hline   
& & & & &\\
$\delta$ & 2 & 3 & 4 & 5 & 6 \\
& & & & &\\
\hline
& & & & &\\                     
$M_D$ (TeV)& 0.57 & 0.36& 0.26 & 0.19 & 0.16\\
& & & & &\\
\hline
\end{tabular} 
\end{center}}
\vspace{2mm}
Table 2: Combined 95\% CL limits on $M_D$ from LEP and Tevatron
data, from ref. \cite{gians}. Notice that $M_D$ here is related to
$M_D$\cite{gians} by $M_D$\cite{gians}$=(2\pi)^{\delta/(2+\delta)} M_D$.
\vspace{3mm}

This signal is theoretically clean. Moreover, as long as
the energy $E$ does not exceed $M_D$, thus remaining
within the domain of validity of the
low-energy effective theory, the predictions do not 
depend on the ultraviolet properties of the theory.

A second type of effect is that induced by virtual KK
graviton exchange. At variance with the previous one,
such effect is sensitive to the ultraviolet physics:
the amplitudes diverge already at the tree level for 
$\delta\ge 2$. In string theory the tree-level amplitudes from
graviton exchange correspond to one-loop amplitudes
for the exchange of open string excitations.
They are compactification dependent and also
sub-leading compared to open string tree-level exchanges \cite{igna}. 
In a more model-independent approach, based on a
low-energy effective field theory, the best that can be done
is to parametrize these effects in terms of effective 
higher-dimensional operators. If we focus on the
sector of the theory including only fermions and gauge bosons, 
there are only two independent
operators up to dimension $d=8$ \cite{chan,gians}:
\be
\frac{c_\tau}{2} \left(T_{\mu\nu}T^{\mu\nu}-
\frac{T_\mu^\mu T_\nu^\nu}{2+\delta}\right)~~~~~~~(d=8)~~~,
\ee 
\be
\frac{c_Y}{2}\left(\sum_f {\bar f} \gamma^\mu \gamma_5 f \right)^2~~~~~~~~~~~~
(d=6)~~~,
\ee
where $T_{\mu\nu}$ is the energy-momentum tensor and
the sum over $f$ extends to all fermions.
The dimension eight operator arises from graviton
exchange at the tree-level. It contributes to dilepton
production at LEP, to diphoton production at Tevatron
and to the scattering $e^\pm p \to e^\pm p$ at Hera.
The present data imply the bound \cite{gians}:
\be
\left(\frac{8}{\vert c_\tau\vert}\right)^{\frac{1}{4}}
> 1.3~{\rm TeV}~~~.
\ee
The dimension six operator arises at one-loop. It is even under charge
conjugation and
singlet under all gauge and global symmetries. It is bounded mainly
from the search of contact interactions at LEP, dijet and Drell-Yan
production at Tevatron and $e^\pm p \to e^\pm p$ at Hera \cite{gians}:
\be
\left(\frac{4 \pi}{\vert c_Y\vert}\right)^{\frac{1}{2}}
> (16\div 21)~{\rm TeV}~~~,
\label{bcy}
\ee
where the two quoted bounds refer, respectively, 
to a negative and a positive $c_Y$.
In terms of the fundamental scale $M_D$ and the cut-off $\Lambda$
of the effective theory, $c_\tau$ and $c_Y$ are expected to scale
as:
\be
c_\tau\approx\frac{\Lambda^{\delta-2}}{M_D^{\delta+2}}~~~~~~~~
c_Y\approx\frac{\Lambda^{2\delta+2}}{M_D^{2\delta+4}}~~~,
\label{estc}
\ee
(more refined estimates can be found in \cite{gians}).
If we naively set $\Lambda\approx M_D$, then the present
limit on $c_Y$ from eq. ref{bcy}
would provide the strongest collider bound on $M_D$.
In a conservative analysis $M_D$ and $\Lambda$ should be kept as
independent. In this case, assuming $M_D\approx 1$ TeV,
we see that the bound (\ref{bcy}) and the estimate (\ref{estc})
imply that $\Lambda$ should be considerably smaller than $M_D$.
Such a situation, where the cut-off scale is required to be much
smaller than the mass scale characterizing the low-energy
effective theory, is not uncommon in particle physics.
For instance the naive estimate of the amplitude for 
$K^0_L\to \mu^+\mu^-$ in the Fermi theory 
gives $G_F^2 \Lambda^2$ and data require $\Lambda\ll 1/\sqrt{G_F}$.
Indeed here the role of $\Lambda$ is played 
by the charm mass, as a consequence of the GIM mechanism.
Similarly, the stringent bound on $c_Y$ could indicate that the modes
needed to cure the ultraviolet behaviour of amplitudes with 
KK graviton exchange are possibly quite light, if
the fundamental scale $M_D$ is close to the TeV range. 

The present sensitivity will be considerably extended
by future colliders, like LHC, that could
probe $M_D$ up to about 3.4(2.3) TeV, for $\delta=2(4)$ \cite{grw}.
The most promising channel is single jet
plus missing transverse energy. Estimates based on the 
low-energy effective theory become questionable for $\delta\ge 5$.
If the energy at future colliders became comparable
to the fundamental scale of gravity $M_D$, 
production and decay of black holes could take place
in our laboratories, with expectations that have been 
review by Landsberg at this conference \cite{land}.
In a even more remote future, collisions at trasplanckian
energies could provide a robust check of these ideas,
especially for the possibility of dealing with gravity
effects in a regime dominated by the classical approximation
\cite{gianr}.

\subsubsection{Limits from astrophysics}
Today the most severe limits on $M_D$ come from astrophysics
and in particular from processes that can influence supernova
formation and the evolution of the daughter neutron star.
There are three relevant processes. The first one
is KK graviton production during the explosion of a supernova,
whose typical temperature of approximately $50$ MeV 
makes kinematically accessible the KK levels for the
compactification scales listed in table 1.
\\[0.1cm]
{\begin{center}
\begin{tabular}{|c|c|c|c|} 
\hline   
& & &\\
& $\delta=2$ & $\delta=3$ &$\delta=4$\\
& & &\\
\hline
& & &\\                         
{\tt SN cooling (SN1987A)} & 8.9 & 0.66 & 0.01\\
& & &\\
\hline
& & &\\                         
{\tt Diffuse gamma rays} & 38.6 & 2.65 & 0.43\\
& & &\\
\hline
& & &\\                         
{\rm NS heat excess} & 701 & 25.5 & 2.77\\
& & &\\
\hline
\end{tabular} 
\end{center}}
\vspace{3mm}
Table 3: Lower bound on $M_D$, in TeV, from astrophysical
processes \cite{han}.

\vskip 0.2cm
The amount of energy carried away by KK gravitons cannot
deplete too much that associated to neutrinos, whose flux   
was observed in SN1987A. A competing process is KK graviton production,
mainly induced by nucleon-nucleon bremsstrahlung 
$N N\to N N {\tt graviton}$, and controlled 
in the low-energy approximation by $M_D$. This process 
should be adequately suppressed. If KK graviton production during
supernova explosions takes place, then a large fraction of gravitons
remains trapped in the neutron star halo
and the subsequent decay of gravitons into photons produces a diffuse
$\gamma$ radiation, which is bounded by the existing measurements
by the EGRET satellite.
The photon flux should also not overheat the neutron star
surface.
The corresponding limit on $M_D$ depends on the assumed 
decay properties of the massive gravitons.
These limits are summarized, for $\delta=2,3,4$ 
in table 3.
These bounds rapidly softens for higher $\delta$, due to
the energy dependence of the relevant cross-sections
in the low-energy approximation. They are strictly related to
the spectrum of KK gravitons, which,
as we can see from table 1, has no sizeable gap compared
to the typical energy of the astrophysical processes 
considered here.

We also recall that KK graviton production can largely affect
the universe evolution \cite{cosmoled}. Going backwards in time, for 
$M_D\approx 1$ TeV, the universe has a standard evolution only 
up to a temperature $T_*$ given approximately by 10 MeV, if
$\delta=2$ and by 10 GeV, if $\delta=3$. For higher temperatures
the production of KK gravitons replaces the
adiabatic expansion as the main source of cooling.
While such low temperatures are still consistent with 
the big-bang nucleosynthesis, they render both inflation
and baryogenesis difficult to implement.

In summary, $M_D\approx 1$ TeV still represents a viable possibility
if $\delta\ge 4$, while for lower $\delta$ and in particular for the special
case $\delta=2$, it is difficult to reconcile the 
expectations of LED, at least in the simplest version
discussed here, with astrophysical data. 

\subsection{Warped compactification}
Both astrophysical and cosmological problems are evaded
if the spectrum of KK gravitons has a sufficient gap.
For instance a gap higher
than the temperature of the hottest present astrophysical
object, approximately 100 MeV, removes all the astrophysical bounds
discussed above.
Such a gap can be obtained in several ways. Up to now
we have assumed an isotropic toroidal compactification.
In general, the KK spectrum depends not only on the
overall volume $V_\delta$, but also on the moduli that
define the shape of the compact space. For particular choices
of these moduli, the KK spectrum displays the desired gap 
\cite{dienes}.

\begin{figure}[h!]
\resizebox{0.30\textwidth}{!}{%
  \includegraphics{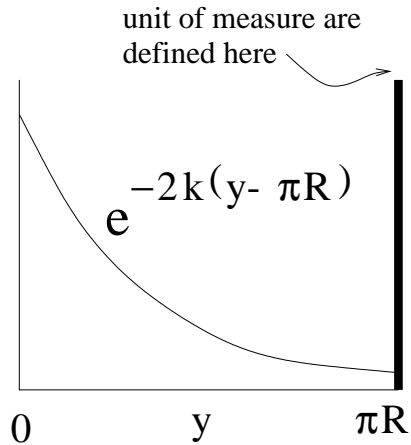}
}
\caption{Dependence of the warping factor on the extra coordinate $y$.
With the adopted parametrization the warping factor is equal to one
at the brane in $y=\pi R$. Mass parameters are measured in units
used by the observer at $y=\pi R$.}
\label{figw}     
\end{figure}

An additional assumption that has been exploited up to
this point is that of a factorized metric 
for the space-time. This assumption is no longer justified
if the underlying geometry admits walls (also referred 
to as branes) carrying some energy density. Then, by
the laws of general relativity, the background metric
is warped and the relation between $M_{Pl}$ and $M_D$
is modified \cite{rs}. For instance, in the Randall-Sundrum set-up,
the metric can be parametrized as:
\be
ds^2=e^{\dd 2 k(y-\pi R)} \eta_{\mu\nu} dx^\mu dx^\nu+dy^2~~~,
\label{rsmetric}
\ee
where $\eta_{\mu\nu}$ is the four-dimensional Minkowski metric and 
$k^{-1}$ is the radius of the AdS space.
Notice that we have rescaled the coordinates $x^\mu$ by
the overall factor $e^{k \pi R}$, compared to the most
popular parametrization of the Randall-Sundrum metric. 
As a result, all mass parameters 
are now measured in units of the typical mass scale at $y=\pi R$, 
the TeV (see fig. \ref{figw}).
The masses $M_{Pl}$ and $M_D$ are now related by:
\be
\left(\frac{M_{Pl}}{M_D}\right)^2 = 
M_D~ \frac{(e^{2 k \pi R}-1)}{k}~~~.
\label{rs}
\ee
This relation can be view, for $\delta=1$ as a generalization
of the one given in eq. (\ref{dtop}) for a factorizable metric.
Indeed, by sending $k$ to zero the metric (\ref{rsmetric})
becomes flat and we recover eq. (\ref{dtop}). 
By comparing eqs. (\ref{rs})
and (\ref{dtop}) we see that the factor multiplying
$M_D$ on the right hand side of eq. (\ref{rs}) plays, in a loose sense, 
the role of the
volume $V_\delta$ of the compact space, 
measured in units used by the observer at $y=\pi R$.
In this case
the dependence of the ``volume'' on the radius $R$ is exponential
\footnote{Such a dependence is also found in other 
geometries of the internal space \cite{kal}.}.
Remarkably, we do not need a huge radius $R$ to achieve 
a large hierarchy
between $M_{Pl}$ and $M_D\approx 1$ TeV. If $k$ and $M_D$
are comparable, then $R\approx 10 k^{-1}$ does the job.
The KK graviton levels, controlled by $1/R$ in the
present parametrization, start now naturally at the TeV
scale. Astrophysical and cosmological bounds do not apply.
Signals at colliders are quite different from those
discussed in the LED case. The KK gravitons have couplings
suppressed by the TeV scale, not by $M_{Pl}$. Their levels
are not uniformly spaced and they are expected to produce
resonance enhancements in Drell-Yan processes. A portion of
the parameter space has already been probed at the Tevatron
collider by CDF, by searching for heavy graviton decays into
dilepton and dijet final states \cite{rscdf}.
In this model the radion mass is expected to be in 
the range $0.1-1$ TeV,
the exact value depending on the specific mechanism 
that stabilizes the radius $R$. It can be produced by gluon
fusion and it mainly decays into a dijet or a $ZZ$ pair,
as the Higgs boson. No significant bounds can be extracted 
from the present radion search.

The Randall-Sundrum setup provides an interesting alternative
to LED. It avoids the tuning of geometrical parameters versus
$M_D$ that is still needed in LED to reproduce 
the hierarchy between $M_{Pl}$ and the electroweak scale.
It overcomes the difficulties related to the presence in LED 
of a ``continuum'' of KK graviton states. 
 
\section{Little Hierarchy Problem}
The assumption that the SM fields are 
confined on a four-dimensional brane that does not
extend into the extra space is too restrictive. 
Strong and electroweak interactions have been successfully
tested only up to energies of order TeV. Therefore a part
of or all the SM fields might have access to
extra dimensions of typical size $L\le({\rm TeV})^{-1}$ (see fig. \ref{fig1}).
There are several theoretical motivations for extra dimensions
at the TeV scale. Already long ago, it was observed that
one way to achieve supersymmetry breaking with
sparticles masses in the TeV range, is through a suitable
TeV compactification \cite{susyb}. Here I will discuss a more recent development,
related to the so-called `little' hierarchy problem.
On the one hand, the present data     
provide an indirect evidence for a gap between the Higgs mass $m_H$, 
required to be small by the precision tests, and the electroweak 
breaking scale, required to be
relatively large by the unsuccessful direct search for new physics.
On the other hand, from the solution of the `big' hierarchy problem,
we would naively expect that a light Higgs boson should require
new light weakly interacting particles (e.g. the 
chargino in the MSSM), that have not been revealed so far.
This gap is not so large and it can be filled either by a moderate
fine-tuning of the parameters in the underlying theory, or by
looking for specific theories where it can be naturally produced.
Extra dimensions at the TeV scale provide in principle a framework
for these more natural theories. Indeed, new weakly interacting
states show up at the TeV scale, whereas the Higgs mass can be
kept lighter by some symmetry.
\subsection{Higgs mass protected by SUSY} 
\begin{figure}[h!]
\resizebox{0.45\textwidth}{!}{%
  \includegraphics{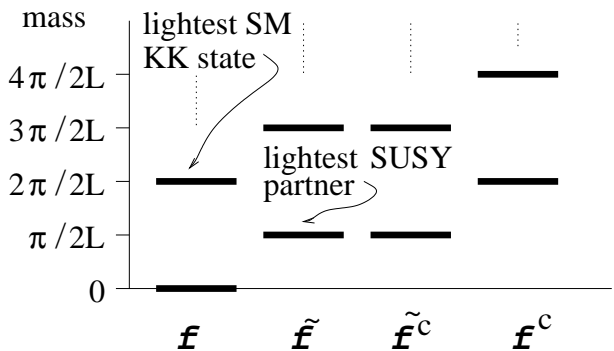}
}
\caption{Mass spectrum of the model in ref. \cite{mbar}
for a generic matter multiplet: 
$f$ and ${\tilde f}$ are respectively
fermion and scalar components of a chiral $N=1$ SUSY multiplet;
$f^c$ and ${\tilde f}^c$ are additional components required by
supersymmetry in D=5.}
\label{figb}     
\end{figure}

In the last years there has been a growing interest in five-dimensional
models where supersymmetry is broken by boundary conditions on an interval
of size $L\approx 1~ {\rm TeV}^{-1}$. Supersymmetry breaking by boundary
condition can reduce the arbitrariness in the soft breaking sector,
thus making the model more predictable \cite{bhn1}. Moreover such a possibility provides 
an alternative to the MSSM with universal boundary condition
at the grand unified scale to study the interplay between
supersymmetry and electroweak symmetry breaking \cite{bhn1,ewsb}.
Also, such a set-up demonstrates very useful to study
important theoretical issues such as the problems of cancellation
of quadratic divergences \cite{quad} and of gauge anomalies 
\cite{anom}.
As in the MSSM, the electroweak symmetry breaking
can be triggered by the top Yukawa coupling. 
For instance, in a particular 
model \cite{mbar} belonging to this class, the Higgs mass is finite and calculable
in terms of two parameters, the length $L$ of the extra dimension 
and a mass parameter $M$ that is responsible for the 
localization of the wave function for the zero mode of the 
top quark.
By including not only leading terms, but also two-loop corrections
originating from the top Yukawa coupling and the strong coupling 
constant, a Higgs mass in the relatively narrow range $m_H=(110\div 125)$
GeV is found, for $L^{-1}=(2\div 4)$ TeV and $2\le L M\le 4$.
The model is characterized by the spectrum of
KK excitations displayed in fig. \ref{figb}.
The KK tower of each ordinary fermion is accompanied by 
a tower for the associated SUSY partner.
The two towers have the same spacing,
$\pi/L$, but they are shifted by $\pi/(2 L)$. 
Two additional towers are required by SUSY in five dimensions.
The detection of this pattern would provide
a distinctive experimental signature of the model.
A peculiar feature of the model is that all the degrees of freedom
are in the bulk (models of this type are said to have
universal extra dimensions \cite{ued}), and the only localized 
interactions are the Yukawa ones. Therefore momentum along the 
fifth dimension is conserved by gauge interactions. 
No single KK mode can be produced through
gauge interactions and no four-fermion operator arises from
tree-level KK gauge boson exchange. 
This property softens the experimental bounds on universal
extra dimensions.

\subsection{Higgs mass protected by gauge symmetry: 
Higgs-gauge unification} 
More than twenty years ago Manton \cite{man} 
suggested that the Higgs
field could be identified with the extra components
of a gauge vector boson living in more than four dimensions.
In this case the same symmetry that protects
gauge vector bosons from acquiring a mass, could help
in preventing large quantum corrections to the Higgs mass
\cite{ms3,prot}.

A simple example of how this idea can be practically implemented
is a Yang-Mills theory in $D>4$ dimensions with gauge group 
SU(3). The gauge vector bosons can be described by a 3$\times$3 
hermitian matrix $A_M$ transforming in the adjoint
representation of SU(3):
\be
A_M=\left(
\begin{array}{c|c}
 & \\
A_M^a & A_M^{\hat a}\\
 & \\
\hline
~~~A_M^{\hat a}~~~ & A_M^a
\end{array}
\right)~~~~~.
\ee
The vector bosons $A_M^a$ $(a=1,2,3,8)$, lying along the diagonal
2$\times$ 2 and 1$\times$1 blocks, are related to the SU(2)$\times$U(1)
diagonal subgroup of SU(3), to be identified with the gauge
group of the SM, while the fields 
$A_M^{\hat a}$ $({\hat a}=4,5,6,7)$, belonging to the off-diagonal
blocks, are instead associated to the remaining generators of SU(3). From
the four-dimensional point of view, $A_M$ describes both
vector bosons $(M[\equiv\mu]=0,1,2,3)$ and scalars 
$(M[\equiv m]=5,...)$. Out of all these degrees of freedom, 
in the low-energy theory we would like to keep
$A_\mu^a$, that have the quantum numbers of
the electroweak gauge vector bosons $\gamma$, $Z$ and $W^\pm$, 
and $A_m^{\hat a}$, transforming exactly as 
complex Higgs doublets $H_m$, under SU(2)$\times$U(1).
On the contrary, the particles related to $A_\mu^{\hat a}$ and $A_m^a$
are unseen and should be somehow eliminated from the low-energy
description. Recalling that all these fields 
have a Fourier expansion containing a zero mode
plus a KK tower, we would like to keep the zero modes
in the desired sector and project away the zero modes for the unseen
states. 
\begin{figure}[h!]
\resizebox{0.45\textwidth}{!}{%
  \includegraphics{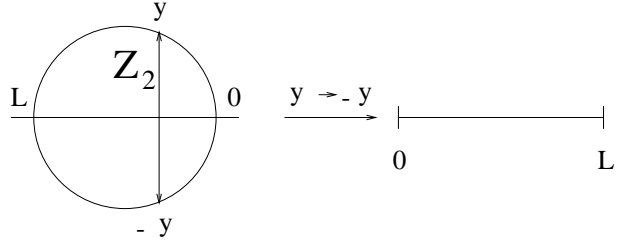}
}
\caption{A $Z_2$ parity symmetry halves a circle $S$ and all the
degrees of freedom originally defined on $S$.}
\label{figz2}     
\end{figure}
Such a projection becomes very natural \cite{prot,ghu} if the compactified
space is an orbifold $S/Z_2$ (see fig. \ref{figz2}), 
where
all the fields are required to have a specific parity under
the inversion of the fifth coordinate: $y\to -y$. It is
sufficient to require that the fields describing the unwanted
states are odd, and all the remaining ones are even.
Such a parity assignment is compatible both with the 
five-dimensional SU(3) gauge symmetry and the five-dimensional 
Lorentz invariance.
To the four-dimensional observer, having access only to the
massless modes, the SU(3) symmetry appears to be broken down to 
SU(2)$\times$U(1). Moreover she/he will count one Higgs doublet $H$ 
for each ED.

In order to promote this fascinating interpretation of the 
gauge-Higgs system to a realistic theory,  
a number of problems should be overcome.  
First, in this example, the request that $H$ has the correct
hypercharge leads to $\sin^2\theta_W=3/4$, a
rather bad starting point. A value of $\sin^2\theta_W(m_Z)$
closer to the experimental result can be obtain by replacing SU(3) with
another group. For instance, the exceptional group $G_2$ leads
to $\sin^2\theta_W=1/4$ \cite{cgm}. 
No large logarithms are expected to modify the
tree-level prediction of $\sin^2\theta_W$, although some
corrections can arise from brane contributions.
Second, electroweak symmetry breaking requires a self-coupling
in the scalar sector. In $D=5$ this coupling can be provided
by D-terms, if the model is supersymmetric \cite{bn}. In $D=6$ the desired
coupling is contained in the kinetic term for the higher dimensional
gauge bosons.
Third, crucial to the whole approach is the absence of
quadratic divergences that could upset the Higgs
lightness. A key feature of these models is a residual
gauge symmetry associated to the broken generators \cite{mar1}. 
It acts on the Higgs field as 
\be
A_5^{\hat a}\to A_5^{\hat a}+\partial_5 
\alpha^{\hat a}
\label{resg}
\ee 
and, in $D=5$ it forbids the occurrence of quadratic
divergences from the gauge vector boson sector. 
In $D=6$ quadratic divergences from the gauge sector 
can be avoided in specific models \cite{mar2}.
Fourth, at first sight it would seem impossible to introduce 
realistic Yukawa couplings, 
given the universality of the Higgs
interactions if only minimal couplings are considered. 
This problem can be solved by allowing for non-local interactions
induced by Wilson lines \cite{cgm,sss}. Left and right-handed fermions are 
introduced at the opposite ends of the extra dimension (see fig.
\ref{figz2}). They feel only the gauge transformations that
do not vanish at $y=0,L$, namely those of SU(2)$\times$U(1) and 
the residual transformation in eq. (\ref{resg}). An interaction
term invariant under both these transformations can be written in term of a
Wilson line:
\be
{\cal L}_{Yuk}=y_{ij}^f~ \overline{f_{Ri}}(0)~ 
P [e^{\dd i\int_0^L dy A_5^{\hat a}(y) T^{\hat a}}]~
f_{Lj}(L)~+h.c.~~~.
\ee
These generic interactions should be further constrained
since the couplings in ${\cal L}_{Yuk}$ may reintroduce
quadratic divergences for $m_H$ at one or two-loop level.

This interesting framework has been largely developed
and improved in the last year. 
It is based on a compactification mechanism able to
provide the electroweak symmetry breaking sector
starting from pure gauge degrees of freedom.
It is characterized by the presence of KK gauge vector 
bosons of an extended group, not necessarily accompanied 
by KK replica for the ordinary fermions.

In a more radical proposal, the electroweak breaking itself
originates directly from compactification,
without the presence of explicit Higgs doublet(s)
\cite{noh}. Higgs-less theories of electroweak interactions
in D=4 are well known. One of their main disadvantages
is that they become strongly interacting at a relatively low
energy scale, around 1 TeV, thus hampering the calculability
of precision observables. In a five-dimensional realization 
an appropriate tower of KK states can delay the violation of
the unitarity bounds beyond about 10 TeV and the low-energy 
theory remains weakly interacting well above the four
dimensional cut-off.
Very recent developments show that such a 
possibility, at least in its present formulation,
it is not compatible with the results of the precision 
tests \cite{nohpt}. 

\subsection{Additional remarks}
\begin{figure}[h!]
\resizebox{0.45\textwidth}{!}{%
  \includegraphics{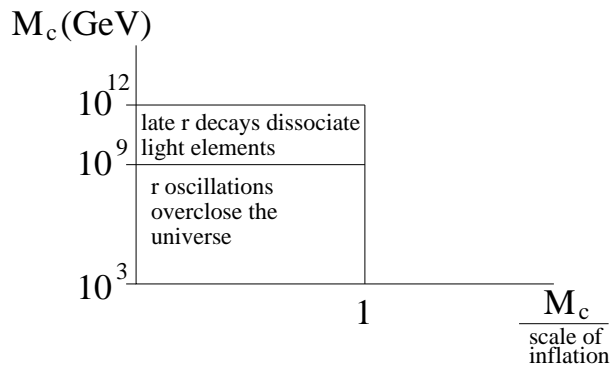}
}
\caption{Cosmological effects of a radion mass of the order
$M_c^2/M_{Pl}$.}
\label{figinfl}     
\end{figure}

Above we have implicitly assumed that some dynamics stabilizes 
the radion VEV at the right scale $M_c\equiv 1/L\approx 1$ TeV.
Radion-matter interactions are controlled by the gravitational
coupling $G_N\propto 1/M_{Pl}^2$. Other radion properties 
depend on the specific stabilization mechanism. 
In several explicit models of weak scale compactification
the radion mass is approximately given by $M_c^2/M_{Pl}\approx 10^{-3}$
eV. This induces calculable deviations from Newton's law at distances 
of the order 100 $\mu$m, 
even in the absence of contributions from KK graviton modes \cite{per}.
Moreover, such a light radion can dangerously modify the property
of the early universe \cite{kol}, if the scale of inflation is larger
than the compactification scale (see fig. \ref{figinfl}).
\vspace{3mm}

There are three general problems that affect theories 
where the presence of ED requires a cut-off scale around 1 TeV. 
These theories include both the LED scenario discussed in section 2
and the TeV compactification we are discussing here.
\vspace{7mm}
\begin{itemize}
\item[$\bullet$]{\bf Conflict with EWPT}
\end{itemize}

\noindent
Potentially large corrections to the electroweak observables
arise when the compactification scale is close to the TeV range,
from the tree-level exchange of KK gauge bosons.
For $\delta>1$ the sum over the KK levels diverges,
a regularization is needed and the related effects 
become sensitive to the ultraviolet completion of the theory.
For $\delta=1$, KK gauge boson exchange and the
mixing between ordinary and KK gauge bosons
typically require to push the compactification scale beyond
few TeV \cite{muck}. In the LED and RS scenarios, these effects are
model dependent and more difficult to estimate \cite{rspt}.

\begin{itemize}
\item[$\bullet$]{\bf B and L approximate conservation}
\end{itemize}

\noindent
Present data suggest that baryon and lepton numbers B and L
are approximately conserved in nature. Stringent experimental
bounds have been set on the proton lifetime. For instance 
\cite{sk}
\be
\tau(p\to e^+ \pi^0)>4.4\times 10^{33}~yr~~~.
\ee
Neutrino masses are extremely small compared to the other
fermion masses
\be
\Delta m^2_{atm}\approx 2\times 10^{-3}~{\rm eV}^2~~~~~
\sum_i m^\nu_i<1~{\rm eV}~~~.
\ee
While there are no fundamental principles requiring exact B/L
conservation and indeed B/L violations are needed by baryogenesis,
these experimental results imply that the amount of B/L breaking
should be tiny in nature.
In a low-energy (SUSY and R-parity invariant) effective theory 
B and L violations are described,
at leading order, by dimension five operators such as
\bea
\int d^2\theta\frac{(H_u L)(H_u L)}{\Lambda}&=&\frac{v^2}{2 \Lambda}\nu\nu+...\nn\\
\int d^2\theta\frac{QQQL}{\Lambda}& &~~~~~~~~~~~~~~.
\label{qqql}
\eea
(Without SUSY the leading B-violating operators have dimension six).
If the cut-off $\Lambda$ 
of the low-energy effective theory is very large,
as the conventional solution of the hierarchy problem via
four-dimensional SUSY seems to indicate, then B and L violating
operators are sufficiently suppressed
(we will reconsider this issue for the baryon number later on).
In a theory characterized by a very low cut-off $\Lambda$, 
as the higher-dimensional theories discussed so far, 
additional mechanisms to deplete
B/L violating operators should be invoked \cite{edbl,split}.

\begin{itemize}
\item[$\bullet$]{\bf Gauge coupling unification}
\end{itemize}

\noindent
One of the few successful experimental indications in favor of
low-energy SUSY is provided by gauge coupling unification.
In the one-loop approximation the prediction for $\alpha_3(m_Z)$
perfectly matches the experimental value \cite{pdg}
\be
\alpha_3(m_Z)=0.117\pm 0.002~~~.
\label{a3exp}
\ee
The inclusion of two-loop effects,
thresholds effects and non-perturbative contributions
raises the theoretical error up to $\delta\alpha_3(m_Z)\approx 0.01$
while maintaining a substantial agreement with data. In $D=4$ gauge coupling unification requires three independent
ingredients: logarithmic running, right content of light particles
and appropriate unification conditions. 
If the ultraviolet cut-off, implied by the presence of extra 
dimensions, is much smaller than the grand unified scale,
then gauge coupling unification becomes a highly non-trivial 
property of the theory.
\vspace{3mm}

It is not possible to review here all the existing proposals
in order to reconcile extra dimensions having a low cut-off
with B and L approximate conservation and gauge coupling 
unification. Summarizing very crudely the state of the art
we can say that these properties are quite natural within
the ultraviolet desert of a (SUSY) four-dimensional low-energy 
theory, especially if such a theory is complemented by
a grand unified picture where the particle classification
is clarified and the required unification condition is
automatic. On the other hand, B and L approximate conservation 
and gauge coupling unification are possible \cite{edbl,split,unif}, 
but far from
generic features if an infrared desert exist: $V_\delta>>M_D^{-\delta}$.
In the absence of a desert, as in the case of compactifications
at the TeV scale, gauge coupling unification is lost
\footnote{The possibility that power-law running \cite{gherg}
enforces gauge coupling unification at a scale close to the TeV 
depends on assumptions about the unknown ultraviolet completion
of the theory \cite{papa}.}.
In the Randall-Sundrum setup the TeV scale is not the highest
accessible energy scale and this leaves open the possibility
of achieving gauge coupling unification in a way 
close to the conventional four-dimensional picture \cite{urs}.

\section{Grand Unification and Extra Dimensions}
The idea of grand unification is extremely appealing.
It automatically provides the unification conditions
for a successful gauge coupling unification.
It sheds light on the otherwise mysterious classification
of matter fields, giving a simple explanation
for electric charge quantization, gauge anomaly cancellation
and suggesting relations among fermion masses.
It provides the ingredients for baryogenesis.
However, in its conventional four-dimensional formulation,
grand unification is affected by two major problems.

The first is the doublet-triplet splitting problem.
In any grand unified theory (GUT) the electroweak 
doublets $H_{u,d}^D$, needed for the electroweak symmetry
breaking, sit in the same representation of the grand
unified group together with color triplets $H_{u,d}^T$.
Doublets are required to be at the electroweak scale,
while triplets should be at or above the unification scale,
to avoid fast proton decay and not to spoil gauge
coupling unification. Such a huge splitting is 
fine-tuned in minimal models, or realized 
naturally at the price of baroque Higgs structures
in non-minimal ones \cite{afm1}. Moreover, even when implemented
at the classical level, it can be upset by radiative corrections
after SUSY breaking and by the presence of non-renormalizable
operators induced by physics at the $M_{Pl}$ scale \cite{racs}.

The second problem is the conflict between the 
proton decay rates, dominated by dimension five operators in 
minimal SUSY GUTs, and experimental data. 
For instance minimal SUSY SU(5) predicts \cite{hisano}:
\be
\tau(p\to K^+ \overline{\nu})\approx 10^{32}
\left(\frac{2 \tan\beta}{1+\tan^2\beta}\right)^2
\left(\frac{m_T({\rm GeV})}{10^{17}}\right)^2
~yr~~~.
\label{thp}
\ee
This prediction is affected by several theoretical
uncertainties (hadronic matrix elements, masses
of supersymmetric particles, additional physical phase
parameters, wrong mass relations between
quarks and leptons of first and second generations) 
but, even by exploiting such uncertainties to stretch
the theoretical prediction up to its upper limit,
the conflict with the present experimental bound
\be
\tau(p\to K^+ \overline{\nu})> 1.9\times 10^{33}~yr~~~(90\%~{\rm CL})~~~,
\ee
is unavoidable (in eq. (\ref{thp}) such uncertainties
have already been optimized to minimize the rate).

\begin{figure}[h!]
\resizebox{0.30\textwidth}{!}{%
  \includegraphics{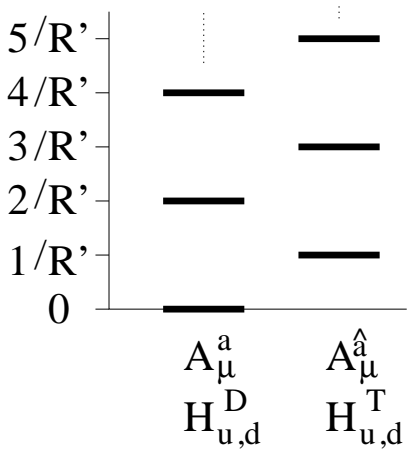}
}
\vskip 0.3cm
\caption{Mass spectrum in the gauge boson and Higgs sectors
of a five-dimensional SU(5) GUT, compactified on $S/(Z_2\times Z_2')$.}
\label{figgut}     
\end{figure}

Remarkably, both these problems can be largely alleviated if the grand
unified symmetry is broken by compactification of a 
(tiny) extra dimension on a orbifold \cite{kawa}.
The gauge symmetry breaking mechanism is exactly the one we 
have already described when discussing gauge-higgs unification.
For instance, in the case of SU(5), 
the gauge vector bosons $A_\mu$ can be assembled in
a 5$\times$5 hermitian traceless matrix as follows:
\be
A_\mu=\left(
\begin{array}{c|c}
 & \\
A_\mu^a & A_\mu^{\hat a}\\
 & \\
\hline
~~~A_\mu^{\hat a}~~~ & A_\mu^a
\end{array}
\right)~~~~~,
\ee
where the diagonal blocks are 3$\times$3 and 2$\times$2 matrices. 
Then $A_\mu^a$ $(a=1,...12)$ are 
the gauge vector bosons of the SM
while $A_\mu^{\hat a}$ $({\hat a}=13,...24)$
are associated to the generators that are in SU(5)
and not in the SM. By working in five dimensions,
the zero modes of $A_\mu^{\hat a}$ can be eliminated
if we require that these fields are odd under the
parity symmetry $Z_2'$: $y'\to -y'$, $y'\equiv y-\pi R'/2$. 
The resulting spectrum is displayed 
in fig. \ref{figgut}.
If we consider a tiny radius of the fifth dimension, 
$1/R'\approx M_{GUT}$ ($M_{GUT}=2.4\times 10^{16}$ GeV 
being the four-dimensional grand unified scale), 
then from the viewpoint of the four-dimensional observer, having access
to energies much smaller than $M_{GUT}$, SU(5) appears to be broken 
down to SU(3)$\times$SU(2)$\times$U(1).
Moreover, if the Higgs multiplets $H_{u,d}$ containing the Higgs doublets
$H_{u,d}^D$ live in the bulk, their $Z_2'$ parity assignment is fixed from the
gauge sector (up to a twofold ambiguity) and the desired DT
splitting can be automatically achieved by compactification \cite{kawa},
as illustrated in fig. \ref{figgut}. 
In the SUSY version
of this model another parity $Z_2$, acting as $y\to -y$, 
removes all the additional states required by SUSY in D=5
and the massless modes are just in $A_\mu^a$ and $H_{u,d}^D$.
The double identification implied by $Z_2\times Z_2'$
reduces the circle $S$ down to the interval $(0,\pi R'/2)$.
The point $y=\pi R'/2$ is special, since all the gauge vector bosons
$A_\mu^{\hat a}$ and the parameters of the corresponding
gauge transformations vanish there. Therefore in $y=\pi R'/2$
the effective gauge symmetry is only SU(3)$\times$SU(2)$\times$U(1),
not the full SU(5).

To complete the solution of the DT splitting problem
and to keep under control proton decay, the mass term
$H_u H_d$ allowed by both SU(5) gauge symmetry and 
$N=1$ five-dimensional supersymmetry \cite{sohnius},
should be adequately suppressed. This can be done by
assuming a U(1)$_R$ symmetry of Peccei-Quinn type, broken 
down to the usual $R$-parity only by
small supersymmetry breaking soft terms \cite{halln}. 
This symmetry keeps the Higgs doublets light 
and, when extended to the matter sector to include 
$R$-parity,
forbids dimension five operators leading to proton decay (see
eq. (\ref{qqql})) and eliminates 
all dangerous B/L violating operators of dimension four.
Alternatively we can assign appropriate parities
to matter fields, to suppress or even to completely forbid
proton decay, at the price of explicitly loosing the 
SU(5) gauge symmetry in the matter sector \cite{af}.

Gauge coupling unification is preserved \cite{halln,gc}. There are several
contributions to the running of gauge coupling constant beyond the 
leading order. As in the conventional
four-dimensional analysis we have two-loop corrections and 
contributions coming from the light thresholds,
the masses of the superpartners of the 
ordinary particles. These corrections raise the
leading order prediction of $\alpha_3(m_Z)$:
\be
\alpha_3(m_Z)=\alpha_3^{LO}(m_Z)
+\delta^{(2)}\alpha_3
+\delta^{(light)}\alpha_3
\approx 0.130~~~.
\label{a3b}
\ee
An additional correction, specific of the present
framework, comes from possible gauge
kinetic terms localized at the brane at $y=\pi R'/2$,
where the effective gauge symmetry is only SU(3)$\times$SU(2)$\times$U(1),
not the full SU(5):
\be
\sum_{i=1,2,3}\frac{1}{g_{bi}^2}\int d^4x~ dy~ \delta(y-\frac{\pi R'}{2})
{F_{\mu\nu}}_i 
F^{\mu\nu}_i~~~~~~.
\label{gb}
\ee
Even if these terms are absent at the classical level,
they are expected to arise from divergent radiative corrections
\cite{hgg}.
Therefore a consistent description of the theory requires
their presence, with $g_{bi}$ as free parameters.
By taking into account both the five-dimensional
gauge kinetic term and the contributions in eq. 
(\ref{gb}), the gauge coupling constants 
$g_i^2$ at the cut-off scale $\Lambda$ are given by
\be
\frac{1}{g_i^2}=\frac{2 \pi R'}{g_5^2}+\frac{1}{g_{bi}^2}~~~.
\label{gl}
\ee
If the SU(5) breaking terms $1/g_{bi}^2$ were similar in size to
the symmetric one, we would loose any predictability.
A predictive framework can be recovered by assuming
that at the scale $\Lambda$ the theory is strongly
coupled. In this case $g_5^2\approx 16\pi^3/\Lambda$,
$g_{bi}^2\approx 16 \pi^2$ and from eq. (\ref{gl}) we estimate
\be
\frac{1}{g_i^2}\approx \frac{\Lambda R'}{8 \pi^2}+
O(\frac{1}{16 \pi^2})~~~.
\label{glp}
\ee
The SU(5) symmetric contribution
dominates over the brane contributions and predicts
gauge couplings of order one, provided $\Lambda R'=O(100)$.
Such a gap between the compactification scale $M_c\equiv 1/R'$
and the cut-off scale $\Lambda$ can in turn independently
affect gauge coupling unification through
heavy threshold corrections. At leading order
these corrections, coming from the particle mass
spectrum around the scale $M_c$, are given by \cite{halln}:
\be
\delta^{(heavy)}\alpha_3=-\frac{3}{7\pi}(\alpha_3^{LO})^2
\log\left(\frac{\Lambda}{M_c}\right)~~~.
\ee   
It is quite remarkable that $\Lambda R'=O(100)$ is precisely
what needed in order to compensate the corrections in eq.
(\ref{a3b}) and bring back $\alpha_3(m_Z)$ to the 
experimental value (\ref{a3exp}). 
By considering the whole set of renormalization
group equations one also finds the preferred values
\be
\Lambda\approx 10^{17}~{\rm GeV}~~~~~
M_c\approx10^{15}~{\rm GeV}~~~.
\ee
The compactification scale 
$M_c\approx10^{15}~{\rm GeV}$ is rather smaller than
$M_{GUT}$ and this greatly affects
the estimate of the proton lifetime.
The presence of non-universal brane kinetic terms,
as given by the strong coupling estimate in eq. (\ref{glp}),
suggests that the theoretical error on the prediction of
$\alpha_3(m_Z)$ is similar to the one affecting the 
four-dimensional SU(5) analysis.

The
fifth dimension has also an interest impact on the description
of the flavour sector. Each matter field can be introduced
either as a bulk field, depending on all the five-dimensional
coordinates, or as a brane field, located at $y=0$ or in 
$y=\pi R'/2$.
Matter fields living in $y=\pi R'/2$,
can in principle be assigned to representations of 
SU(3)$\times$SU(2)$\times$U(1) that do not form complete
multiplets under SU(5). For instance we could replace 
the Higgs multiplets, so far regarded as bulk fields,
with brane four-dimensional fields localized in $y=\pi R'/2$.
In this case we could avoid the inclusion of the colour triplet
components, limiting ourselves to the SU(2) doublets alone.
This possibility would provide a radical solution to
the DT splitting problem \cite{hmr}, which cannot be contemplated
in the four-dimensional construction. 

To maintain the power of SU(5) in particle classification, it is preferable 
to introduce fermions and their supersymmetric partners as 
bulk fields or as brane fields localized at $y=0$, where
the full SU(5) symmetry is effective. Zero modes from fermion bulk fields
differs from brane fermions in two respects. First,
each bulk field experiences the same splitting that characterizes
the gauge and the Higgs multiplets (see fig. \ref{figgut}).
Therefore to produce the zero modes of a complete SU(5) representation
two identical bulk multiplets, with opposite $Z_2'$
parities are needed. Second the zero modes of bulk fields
have a $y$-constant wave function carrying the characteristic
suppression $\epsilon\equiv1/\sqrt{\Lambda R}$ (see eq. (\ref{fourier})),
which, as we have seen before,
is of the same order of the Cabibbo angle. 
As a consequence, Yukawa couplings between zero modes arising
from bulk fields are depleted with respect to those
between brane fields, the relative suppression factor
being $\epsilon^2$, $\epsilon$ and  1, respectively, for
bulk-bulk, bulk-brane, brane-brane interactions. 
Moreover only brane-brane Yukawa interactions can lead
to the SU(5) mass relation $m_e=m_d$, since the doubling
of SU(5) representations for bulk matter fields lead
to SU(5)-unconstrained couplings between the zero modes. 
All this suggest to localize the third generation on
the $y=0$ brane, while choosing bulk fields for at least
a part of the matter in the first and second generation.  
In this way the successful relation $m_b=m_\tau$ of
minimal SU(5) is maintained, while the unwanted analogous
relations for the first two generations are lost.

The most relevant signature of GUTs is represented
by proton decay. In the five-dimensional SU(5) model
under discussion, this process is dominated by the exchange 
of the gauge vector bosons $A_\mu^{\hat a}$ \cite{halln,ptd5}. In minimal SUSY
SU(5) such a contribution is controlled by the unification scale
$M_{GUT}\approx 10^{16}$ GeV and, by itself, would give rise to 
a proton life of the order $10^{36}~ yr$, too long to be observed
at present and foreseen facilities. On the contrary, in the 
five-dimensional SU(5) realization, the masses of the lightest gauge vector bosons 
$A_\mu^{\hat a}$ are at the compactification scale 
$M_c\approx 10^{15}$ GeV, which means an enhancement of four
order of magnitudes in the proton decay rate.
Such a huge enhancement is in part balanced by suppression
factors coming either from the mixing angles needed
to relate the third generation living at $y=0$ to the
lightest generations, or from non-minimal brane couplings
between $A_\mu^{\hat a}$ and light bulk fermions.
These suppression factors are also the main source of the large
uncertainty in the estimate of the proton lifetime.
On the other hand, all the uncertainties coming from the 
supersymmetry breaking sector of the theory, which affect
$p$-decay dominated by the triplet higgsino exchange,
are absent. The proton lifetime is expected to be close to
$10^{34}~ yr$ and the main decay channels are $e^+ \pi^0$,
$\mu^+ \pi^0$, $e^+ K^0$, $\mu^+ K^0$, ${\bar \nu} \pi^+$,
${\bar \nu} K^+$. 

This framework has also been extended to larger grand unified groups
like SO(10) and E$_6$ \cite{so10}. The gauge symmetry breaking
of the GUT symmetry down to the SM one is accomplished 
partly by the compactification mechanism that, in its simplest
realization, does not lower the rank of the group and 
requires more than one ED and partly by a conventional 
Higgs mechanism.

\section{Flavour problem}
A realistic description of fermion masses in a four dimensional
framework typically requires either a large number of parameters
or a high degree of complexity and we are probably unable to 
select the best model among the very many existing ones. Moreover,
in four dimensions we have little hopes to understand why
there are exactly three generations. These difficulties might
indicate that at the energy scale characterizing
flavour physics a four-dimensional description breaks down.
This happens in superstring theories.
In the ten-dimensional heterotic string six dimensions
can be compactified on a Calabi-Yau manifold \cite{can} or on orbifolds
\cite{dix} and the flavour properties are strictly related to the
features of the compact space. In Calabi-Yau compactifications
the number of chiral generations is proportional to the Euler
characteristics of the manifold. In orbifold compactifications,
matter in the twisted sector is localized around the orbifold
fixed points and their Yukawa couplings, arising from world-sheet
instantons, have a natural geometrical interpretation \cite{orb}.
Recently string realizations where the light matter fields 
of the SM
arise from intersecting branes have been proposed. 
Also in this context the flavour dynamics is controlled by
topological properties of the geometrical construction \cite{int}, having no 
counterpart in four-dimensional field theories.

It has soon been realized that also in a field theoretical description
the existence of extra dimensions could have important consequences
for the flavour problem. For instance in orbifold compactifications
light four-dimensional fermions may be either localized at the orbifold
fixed points or they may arise as zero modes of higher-dimensional 
spinors, with a wave function suppressed by the square root of the 
volume of the compact space (see eq. (\ref{fourier})). 
This led to several interesting proposals.
For instance, as already discussed in section 2.1.2, we can describe 
neutrino masses by allowing right-handed
sterile neutrinos to live in the bulk of a large fifth dimension \cite{ednu}.
We have also seen that in five-dimensional grand unified theories 
the heaviness of the third generation can be explained by localizing 
the corresponding
fields on a fixed point, whereas the relative lightness of the
first two generations as well as the breaking of the unwanted
mass relations can be obtained by using bulk fields \cite{halln,gutfl}.

\begin{figure}[h!]
\resizebox{0.48\textwidth}{!}{%
  \includegraphics{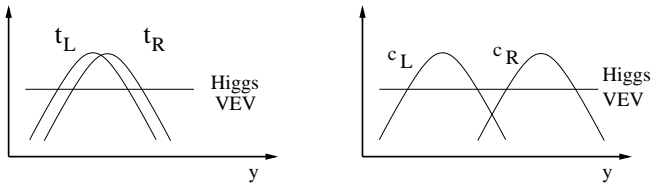}
}
\caption{A different relative localization of left and right-handed
wave functions for top and charm quarks
produces different overlaps with a constant Higgs VEV.}
\label{figze1}     
\end{figure}

\begin{figure}[h!]
\resizebox{0.48\textwidth}{!}{%
  \includegraphics{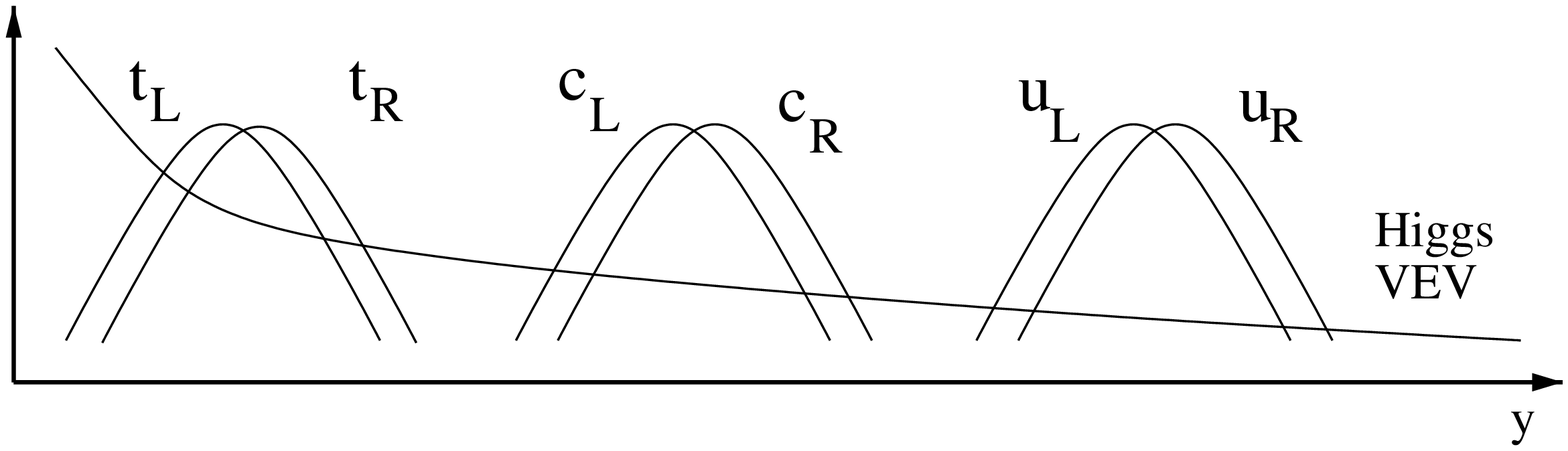}
}
\caption{An equal relative localization of left and right-handed
wave functions for up-type quarks
produces different overlaps with a non-uniform Higgs VEV.}
\label{figze2}     
\end{figure}

Even more interesting is the case when a higher dimensional spinor 
interacts with a non-trivial background of solitonic type. 
It has been known for a long time that this provides a mechanism to 
obtain massless four-dimensional chiral fermions \cite{sol,sol2}. 
For instance, the four-dimensional zero modes of a five-dimensional fermion
$(\psi_L,\psi_R)$ interacting with a real scalar background $\varphi(x_5)$
are formally given by
\be
\psi_{L,R}(x,x_5)\propto e^{\dd\pm g\int_{x_5^0}^{x_5} du~\varphi(u)}
\psi_{L,R}(x)~~~.
\label{zm}
\ee
If $\varphi(x_5)$ is a soliton, with 
$\varphi(\pm\infty)=\pm{\varphi_\infty}$ $({\varphi_\infty}>0)$
only one of the two solutions in eq. (\ref{zm})
is normalizable: $\psi_L$ $(\psi_R)$ if $g>0$ $(g<0)$.
Moreover, since the wave function is localized 
around the core $x_5=x_5^0$ of the topological defect, where 
$\varphi(x_5^0)=0$,
such a mechanism can play a relevant role in explaining the observed 
hierarchy in the fermion spectrum \cite{split}.
Mass terms arise dynamically from the
overlap among fermion and Higgs wave functions (see figs. \ref{figze1}
and \ref{figze2}).
Typically, there is an exponential mapping between the parameters
of the higher-dimensional theory and the four-dimensional masses
and mixing angles, so that even with parameters of order one
large hierarchies are created \cite{ark2}. 
In orbifold compactifications, solitons are simulated by
scalar fields with a non-trivial parity assignment
that forbids constant non-vanishing VEVs.
Also in this case the zero modes of the Dirac operator in such a 
background can be chiral and localized in specific regions of
the compact space. 

A quite interesting possibility arising in models of this sort, 
is that several zero modes can originate from 
a single higher-dimensional spinor \cite{sol,sol2}, 
thus providing an elegant mechanism for understanding the fermion replica.
For instance, in the
model studied in ref. \cite{russi} there is a vortex solution that arises
in the presence of two infinite extra dimensions.     
It is possible to choose the vortex background
in such a way that the number of chiral zero modes of the 
four-dimensional Dirac operator is three.
Each single six-dimensional spinor gives rise to three massless
four-dimensional modes with the same quantum numbers.
Recently this model has been extended to the case of compact extra
dimensions \cite{russi2}. 

In orbifold compactifications similar results can be obtained.
Matter can be described by vector-like D-dimensional fermions
with the gauge quantum numbers of one SM generation.
As a result, the model has neither bulk nor localized gauge anomalies. 
The different generations arise as zero modes of the four-dimensional 
Dirac operator 
by eliminating the unwanted chiralities of the 
D-dimensional spinors through an orbifold projection.
By consistency, D-dimensional fermion masses are required to 
transform non-trivially under the discrete symmetry defining the
orbifold and, as a consequence, the independent zero modes are
localized in different regions of the extra space.
If the Higgs VEV is not constant in the extra space,
but concentrated around some particular point, fermion masses
will acquire the desired hierarchy (see fig. \ref{figze2}).
A toy model successfully implementing this program in the
case of two fermion generations has been recently build 
\cite{bfmpv},
by working with two extra dimensions compactified 
on a orbifold $T^2/Z_2$. In this model
a non-trivial flavour mixing is related to a soft breaking 
of the six dimensional parity symmetry.
In particular, 
the empirical relation $\theta_C\approx \sqrt{m_d/m_s}$ can
be easily accommodated.

\begin{figure}[h!]
\resizebox{0.40\textwidth}{!}{%
  \includegraphics{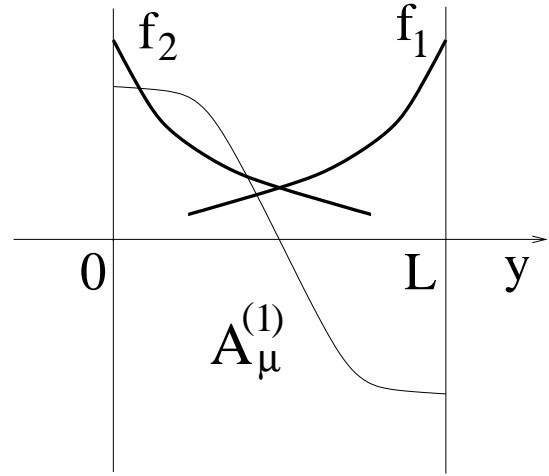}
}
\caption{The non-constant wave function of the KK gauge vector boson
$A_\mu^{(1)}$ can give rise to different coupling to first ($f_1$) and second
($f_2$) fermion generations.}
\label{figze3}     
\end{figure}

The possibility of testing experimentally this idea is strictly related to
the typical size of the extra dimension involved.
When fermions of different generations have 
wave functions with different profiles along the extra dimensions, 
new sources of flavour violation appear \cite{fcnced}. 
Higher KK gauge boson excitations have non-constant wave functions
and this gives rise to non-universal interactions with ordinary fermions
in four dimensions (see fig. \ref{figze3}). 
Exchange of KK gauge bosons produce 
four-fermion interactions, which, after rotation from flavour to mass 
eigenstate basis, mediate flavour-changing
neutral currents. The current limits on FCNC can probe 
compactification scales up to about 100 TeV. 

\section{Cosmological constant problem}
To a good approximation, our four-dimensional space-time possesses an approximate 
Poincar\'e symmetry and, by the law of general relativity, 
this is attributed to a tiny vacuum energy density
$\Lambda_{CC}$ of our universe. 
There are two puzzling aspects of such
an interpretation. First, from the knowledge
of the mass scales implied in fundamental interactions 
we would estimate typical values 
of $\Lambda_{CC}$ that are 
many order of magnitudes bigger than the ``observed'' one. 
Second, we do not understand why the vacuum energy density
is of the same order of magnitude of the matter energy density
today. We conclude this review 
with a qualitative comment about the relevance of extra 
dimensions to the first aspect of the problem.

In four space-time dimensions it is not possible to achieve
a natural cancellation of the cosmological
constant (which would represent already a good starting
point to understand its smallness)
\cite{wein}. Very roughly, in four dimensions
the requirement of general covariance imply the existence
of a massless graviton, universally coupled to all sources.
Therefore, independently from the size of the source, the
gravitational coupling is always given by the Newton constant
$G_N=1/(8\pi M_{Pl}^2)$. The laws of general relativity
demand that the vacuum energy of the universe curves our space-time
by the amount:
\be
H^2\propto G_N \Lambda_{CC}\propto \frac{\Lambda_{CC}}{M_{Pl}^2}
\ee
Notice that the present measurements are only sensitive to
the left-hand side, and the tiny value $\Lambda_{CC}\approx
(10^{-3} {\rm eV})^4$ is 
a consequence of the universal graviton coupling, which
in four dimensions cannot be questioned.

In more than four space-time dimensions this conclusion
is not inescapable since, under certain conditions,
$\Lambda_{CC}$ can curve the extra space leaving our space-time
essentially flat \cite{rsd}. For instance, this is what happens, 
at the price of a fine-tuning, in the Randall-Sundrum model 
described in section 2.2. If these conditions could be 
naturally enforced, this would allow to reconcile a large 
vacuum energy density $\Lambda_{CC}$ with the observed smallness 
of the four-dimensional space-time curvature $H$.
The four-dimensional observer would interpret the absence
of space-time curvature as a modification of gravity 
at very large distances, with a substantial reduction of 
the gravitational coupling to the vacuum energy density. 
Such a mechanism might take place in the presence of
an effective gravitational coupling $G_N(\lambda)$ depending on the 
wavelength of the source. For instance, for wavelengths smaller 
that the present Hubble distance, $G_N(\lambda)$ 
could coincide with the Newton constant
\be
G_N(\lambda)=\frac{1}{8 \pi M_{Pl}^2}~~~~~~~~~~~
\lambda\le H_0^{-1}\approx 10^{28}~{\rm cm}~~~,
\ee
not to induce deviations from the standard cosmology.
For larger wavelengths, the gravitational coupling could be
much smaller:
\be
G_N(\lambda)\ll \frac{1}{8 \pi M_{Pl}^2}~~~~~~~~~~
\lambda> H_0^{-1}\approx 10^{28}~{\rm cm}~~~.
\ee
Since the vacuum energy represents the source with the largest
wavelength $\lambda\gg H_0^{-1}$, a large $\Lambda_{CC}$ could
induce a relatively small four-dimensional space-time curvature,
compatible with the present observations.

Until recently there were no explicit examples of consistent
theories where the behaviour of gravity is modified at large distances.
A substantial progress is represented by models
where the SM fields are localized on a brane in infinite 
volume ED \cite{dgp}. 
In these models gravity along the brane changes from
a four-dimensional regime at small distances to a higher-dimensional 
regime at very large distances and this gives rise to
an effective gravitational coupling $G_N(\lambda)$ with the kind 
of dependence described above \cite{dgs}.

The existence of a large hierarchy between two mass scales is
not avoided in these models. In particular, to make the cross-over
distance sufficiently large, the scale of gravity in the higher-dimensional theory should be quite small, of the order of $10^{-3}$ eV.
This hierarchy can however be made technically stable. 
An interesting feature is represented by expected modifications
of the Newton's law at distances below 1 mm,
an aspect that is also common to other approaches to the cosmological
constant problem \cite{sundrum}, where point-like gravity breaks down
around the 100 $\mu$m scale.
The physical implications of this class of models are presently
under investigation and it is not clear if the difficulties
related to the effective low-energy description \cite{diff} can be
overcome by searching for an embedding in the context
of a fundamental theory such as string theory \cite{emb}. 
It is nevertheless interesting to have  
concrete examples where non-local modifications
of gravity, consistent with the equivalent principle, can be 
analyzed \cite{addg}.

\section*{Acknowledgements}
I am grateful to Guido Altarelli, Carla Biggio, Andrea Brignole, 
Antonio Masiero, 
Isabella Masina, Manuel Perez-Victoria, Riccardo Rattazzi, Tony Riotto and
Fabio Zwirner for useful discussions while preparing this talk.

%
% BibTeX users please use
% \bibliographystyle{}
% \bibliography{}

\end{document}